\documentclass[aip,jcp,amsmath,amssymb,preprint]{revtex4-1}

\usepackage{dcolumn} 
\usepackage{bm}
\usepackage{graphicx}
\usepackage{longtable}
\usepackage{footnote}
\usepackage{float}

\usepackage[version=3]{mhchem}

\usepackage{color}

\draft 

\begin{document}

\newcommand{\bra}[1]{\langle #1|}
\newcommand{\ket}[1]{|#1\rangle}
\newcommand{\braket}[2]{\langle #1|#2\rangle}
\newcommand{\p}{^\prime}
\newcommand{\pp}{^{\prime\prime}}

\title{The rotation-vibration spectrum of methyl fluoride from first principles}

\author{Alec Owens} 
 \email{alec.owens@cfel.de}
\affiliation{Center for Free-Electron Laser Science (CFEL), Deutsches Elektronen-Synchrotron DESY, Notkestrasse 85, 22607 Hamburg, Germany}
\affiliation{The Hamburg Center for Ultrafast Imaging, Universit\"{a}t Hamburg, Luruper Chaussee 149, 22761 Hamburg, Germany}

\author{Andrey Yachmenev}
\affiliation{Center for Free-Electron Laser Science (CFEL), Deutsches Elektronen-Synchrotron DESY, Notkestrasse 85, 22607 Hamburg, Germany}
\affiliation{The Hamburg Center for Ultrafast Imaging, Universit\"{a}t Hamburg, Luruper Chaussee 149, 22761 Hamburg, Germany}

\author{Jochen K\"upper}
\affiliation{Center for Free-Electron Laser Science (CFEL), Deutsches Elektronen-Synchrotron DESY, Notkestrasse 85, 22607 Hamburg, Germany}
\affiliation{The Hamburg Center for Ultrafast Imaging, Universit\"{a}t Hamburg, Luruper Chaussee 149, 22761 Hamburg, Germany}
\affiliation{Department of Physics, Universit\"{a}t Hamburg, Luruper Chaussee 149, 22761 Hamburg, Germany}

\author{Sergei N. Yurchenko}
\affiliation{Department of Physics and Astronomy, University College London, Gower Street, WC1E 6BT London, United Kingdom}

\author{Walter Thiel}
\affiliation{Max-Planck-Institut f\"{u}r Kohlenforschung, Kaiser-Wilhelm-Platz 1, 45470 M\"{u}lheim an der Ruhr, Germany}

\date{\today}

\begin{abstract}
Accurate \textit{ab initio} calculations on the rotation-vibration spectrum of methyl fluoride (\ce{CH3F}) are reported. A new nine-dimensional potential energy surface (PES) and dipole moment surface (DMS) have been generated using high-level electronic structure methods. Notably, the PES was constructed from explicitly correlated coupled cluster calculations with extrapolation to the complete basis set limit and considered additional energy corrections to account for core-valence electron correlation, higher-order coupled cluster terms beyond perturbative triples, scalar relativistic effects, and the diagonal Born-Oppenheimer correction. The PES and DMS are evaluated through robust variational nuclear motion computations of pure rotational and vibrational energy levels, the equilibrium geometry of \ce{CH3F}, vibrational transition moments, absolute line intensities of the $\nu_6$ band, and the rotation-vibration spectrum up to $J=40$. The computed results show excellent agreement with a range of experimental sources, in particular the six fundamentals are reproduced with a root-mean-square error of 0.69~cm$^{-1}$. This work represents the most accurate theoretical treatment of the rovibrational spectrum of \ce{CH3F} to date.
\end{abstract}


\maketitle 

\section{Introduction}

Methyl fluoride (\ce{CH3F}) was one of the first five-atom systems to be treated by full-dimensional variational calculations~\cite{Dunn:JCP86:5088} after a period of pioneering studies on polyatomic molecules.~\cite{Carney:ACP86:305} The field has since gone from strength to strength and accurate rotation-vibration computations on small molecules are nowadays fairly routine.~\cite{Tennyson:JCP145:120901} This has enabled a range of applications such as the production of comprehensive molecular line lists to model hot astronomical objects,~\citep{Tennyson:MNRAS425:21,Tennyson:JMS327:73,Tennyson:IJQC117:92} to probing fundamental physics and a possible space-time variation of the proton-to-electron mass ratio.~\cite{Owens:MNRAS450:3191,Owens:MNRAS454:2292,Owens:PRA93:052506,Owens:MNRAS473:4986}

 The starting point of any variational calculation is the potential energy surface (PES) and its quality will largely dictate the accuracy of the predicted rovibrational energy levels, and to a lesser extent, the transition intensities. Thanks to sustained developments in electronic structure theory it is now possible to compute vibrational energy levels within $\pm$1~cm$^{-1}$ from a purely \textit{ab initio} PES.~\citep{Polyansky:Science299:539,Schwenke:SpecActA58:849,Yachmenev:JCP135:074302,Malyszek:JCC34:337,Owens:JCP142:244306,Owens:JCP143:244317,Owens:JCP145:104305} To do so requires the use of a one-particle basis set near the complete basis set (CBS) limit and the treatment of additional, higher-level (HL) corrections to recover more of the electron correlation energy.~\cite{Helgaker:MP106:2107,Peterson:TCA131:1079} Similarly, transition intensities from first principles, which requires knowledge of the molecular dipole moment surface (DMS), are now comparable to, if not more reliable in some instances, than experiment.~\cite{Yurchenko:CM:183,Tennyson:JMolSpec298:1}

 Although the rovibrational spectrum of \ce{CH3F} has been well documented, its theoretical description is not reflective of the current state-of-the-art in variational calculations. Notable recent works include the PESs and energy level computations of Manson {\it et al.}~\cite{Manson:PCCP8:2848,Manson:PCCP8:2855}, \citet{Nikitin:JMolSpec274:28} and Zhao {\it et al.}~\cite{Zhao:JCP144:204302,Zhao:JCP148:074113}. Theoretical \ce{CH3F} spectra are also available from the TheoReTS database~\citep{Rey:JMolSpec327:138} for a temperature range of 70--300~K but details on the calculations are unpublished except for the PES.~\citep{Nikitin:JMolSpec274:28} In this work, high-level \textit{ab initio} theory is used to generate a new PES and DMS for \ce{CH3F}. The surfaces are represented by suitable symmetrized analytic representations and then evaluated through robust variational nuclear motion calculations. Computed results are compared against a variety of experimental spectroscopic data to provide a reliable assessment of our theoretical approach.

  The paper is structured as follows: In Sec.~\ref{sec:PES}, the electronic structure calculations and analytic representation of the PES are presented. Likewise, the details of the DMS are given in Sec.~\ref{sec:DMS}. The variational calculations are described in Sec.~\ref{sec:variational}. Our theoretical approach is then assessed in Sec.~\ref{sec:results} where we compute the equilibrium geometry of \ce{CH3F}, pure rotational energies, vibrational $J=0$ energy levels, vibrational transition moments, absolute line intensities of the $\nu_6$ band, and the rotation-vibration spectrum up to $J=40$. Concluding remarks are offered in Sec.~\ref{sec:conc}.

\section{Potential energy surface}
\label{sec:PES}

\subsection{Electronic structure calculations}

 Similar to our previous work on \ce{SiH4}~\cite{Owens:JCP143:244317} and \ce{CH4},~\citep{Owens:JCP145:104305} the goal is to construct a PES which possesses the ``correct'' shape. Obtaining tightly converged HL energy corrections with respect to basis set size is less important. Using a focal-point approach,~\citep{Csaszar:JCP108:9751} the total electronic energy was represented as
\begin{equation}\label{eq:tot_en}
E_{\mathrm{tot}} = E_{\mathrm{CBS}}+\Delta E_{\mathrm{CV}}+\Delta E_{\mathrm{HO}}+\Delta E_{\mathrm{SR}}+\Delta E_{\mathrm{DBOC}} .
\end{equation}
The energy at the CBS limit $E_{\mathrm{CBS}}$ was computed with the explicitly correlated F12 coupled cluster method~\cite{Adler:JCP127:221106} CCSD(T)-F12b in conjunction with the F12-optimized correlation consistent polarized valence basis sets, cc-pVTZ-F12 and cc-pVQZ-F12.~\citep{Peterson:JCP128:084102} Calculations employed the frozen core approximation with the diagonal fixed amplitude ansatz 3C(FIX)~\citep{TenNo:CPL398:56} and a Slater geminal exponent value of $\beta=1.0$~$a_0^{-1}$.~\citep{Hill:JCP131:194105} The OptRI,~\citep{Yousaf:JCP129:184108} cc-pV5Z/JKFIT~\citep{Weigend:PCCP4:4285} and aug-cc-pwCV5Z/MP2FIT~\citep{Hattig:PCCP7:59} auxiliary basis sets (ABS) were used for the resolution of the identity and the two density fitting basis sets, respectively. Unless stated otherwise calculations were performed with MOLPRO2012.~\citep{Werner:WCMS2:242}

 A parameterized, two-point formula~\citep{Hill:JCP131:194105} was chosen to extrapolate to the CBS limit,
\begin{equation}\label{eq:cbs_extrap}
E^{C}_{\mathrm{CBS}} = (E_{n+1} - E_{n})F^{C}_{n+1} + E_{n} .
\end{equation}
The coefficient $F^{C}_{n+1}$ is specific to the CCSD-F12b or (T) component of the total CCSD(T)-F12b energy and values of $F^{\mathrm{CCSD-F12b}}=1.363388$ and $F^{\mathrm{(T)}}=1.769474$ were chosen.~\cite{Hill:JCP131:194105} The Hartree-Fock (HF) energy was not extrapolated, rather the HF+CABS (complementary auxiliary basis set) singles correction~\citep{Adler:JCP127:221106} computed in the larger basis set was used.

 The contribution from core-valence (CV) electron correlation $\Delta E_{\mathrm{CV}}$ was determined using the CCSD(T)-F12b method with the F12-optimized correlation consistent core-valence basis set, cc-pCVTZ-F12.~\citep{Hill:JCP132:054108} The same ansatz and ABS as in the $E_{\mathrm{CBS}}$ calculations were employed, however, the Slater geminal exponent was set to $\beta=1.4$~$a_0^{-1}$.
 
 To account for higher-order (HO) correlation we employed the hierarchy of coupled cluster methods such that $\Delta E_{\mathrm{HO}} = \Delta E_{\mathrm{T}} + \Delta E_{\mathrm{(Q)}}$. The full triples contribution $\Delta E_{\mathrm{T}} = E_{\mathrm{CCSDT}}-E_{\mathrm{CCSD(T)}}$, and the perturbative quadruples contribution $\Delta E_{\mathrm{(Q)}} = E_{\mathrm{CCSDT(Q)}}-E_{\mathrm{CCSDT}}$. Calculations with the CCSD(T), CCSDT, and CCSDT(Q) methods were carried out in the frozen core approximation using the general coupled cluster approach~\citep{Kallay:JCP123:214105,Kallay:JCP129:144101} as implemented in the MRCC code~\cite{MRCC:2017} interfaced to CFOUR.~\citep{CFOUR:2017} The full triples and perturbative quadruples utilized the correlation consistent cc-pVTZ and cc-pVDZ basis sets,~\citep{Dunning:JCP90:1007} respectively.
 
 Scalar relativistic (SR) effects $\Delta E_{\mathrm{SR}}$ were obtained using the second-order Douglas-Kroll-Hess approach~\citep{Douglas:AP82:89,Hess:PRA33:3742} at the CCSD(T)/cc-pVQZ-DK~\citep{deJong:JCP114:48} level of theory in the frozen core approximation. The spin-orbit interaction was not considered for the present study as this can be ignored for light, closed-shell molecules in spectroscopic calculations.~\citep{Tarczay:MP99:1769}
 
 The diagonal Born-Oppenheimer correction (DBOC) $\Delta E_{\mathrm{DBOC}}$ was computed with all electrons correlated using the CCSD method~\citep{Gauss:JCP125:144111} implemented in CFOUR with the aug-cc-pCVDZ basis set. Because the DBOC is mass dependent its inclusion means the PES is only applicable for \ce{^{12}CH3F}.  
 
 Grid points were generated randomly using an energy-weighted sampling algorithm of Monte Carlo type. The global grid was built in terms of nine internal coordinates: the C{--}F bond length $r_0$; three C{--}H bond lengths $r_1$, $r_2$ and $r_3$; three $\angle(\mathrm{H}_i\mathrm{CF})$ interbond angles $\beta_1$, $\beta_2$ and $\beta_3$; and two dihedral angles $\tau_{12}$ and $\tau_{13}$ between adjacent planes containing H$_i$CF and H$_j$CF. All terms in Eq.~\eqref{eq:tot_en} were calculated on a grid of 82,653 points with energies up to $h c \cdot 50{\,}000$~cm$^{-1}$ ($h$ is the Planck constant and $c$ is the speed of light) and included geometries in the range $1.005\leq r_0 \leq 2.555$~{\AA}, $0.705\leq r_i \leq 2.695$~{\AA}, $45.5\leq \beta_i \leq 169.5^{\circ}$ for $i=1,2,3$ and $40.5\leq \tau_{jk} \leq 189.5^{\circ}$ with $jk=12,13$. 

\begin{figure}[ht]
\centering
\includegraphics[width=\textwidth]{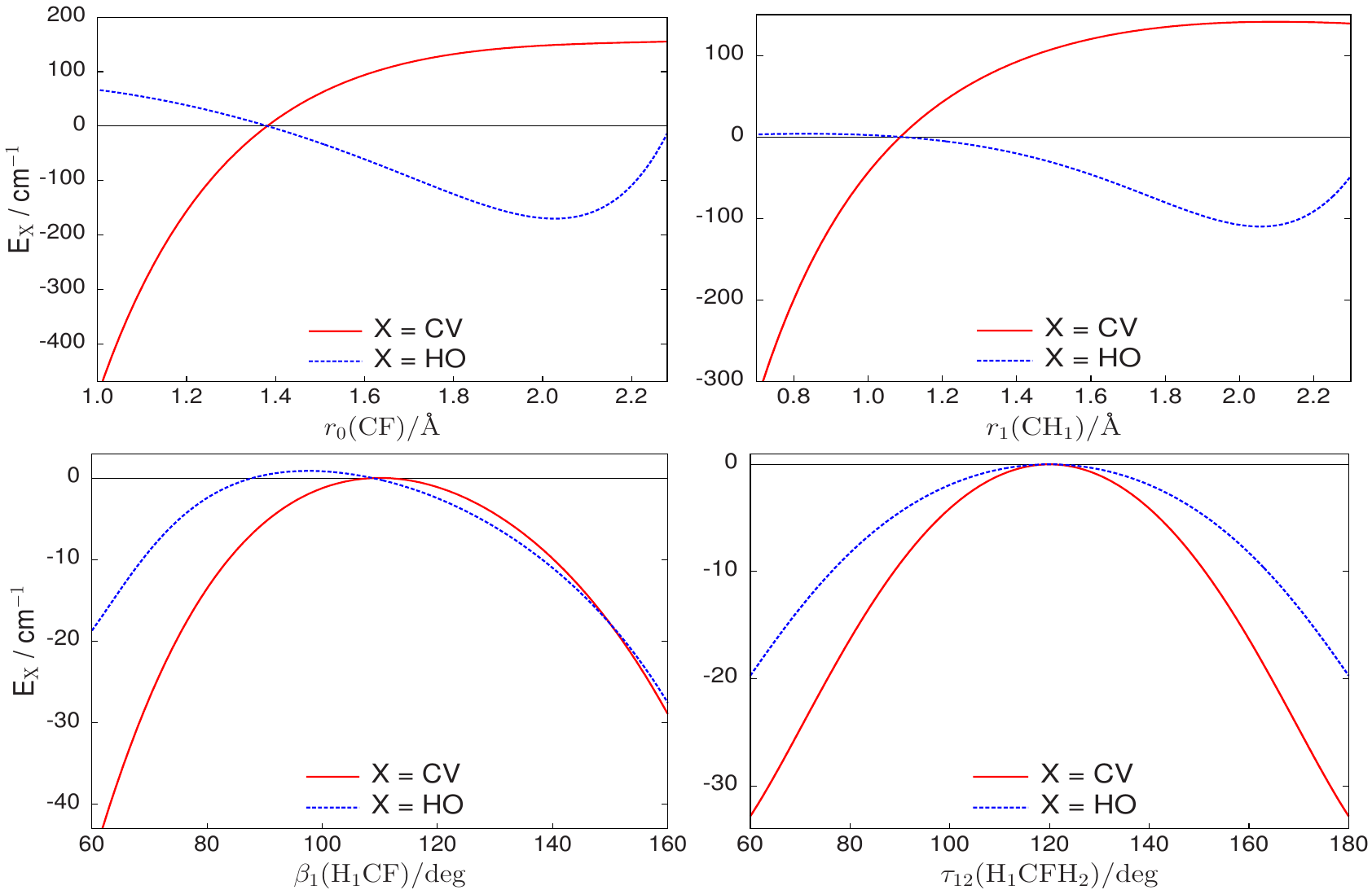}
\caption{\label{fig:1d_cv_ho}One-dimensional cuts of the core-valence (CV) and higher-order (HO) corrections with all other coordinates held at their equilibrium values.}
\end{figure}

\begin{figure}[ht]
\centering
\includegraphics[width=\textwidth]{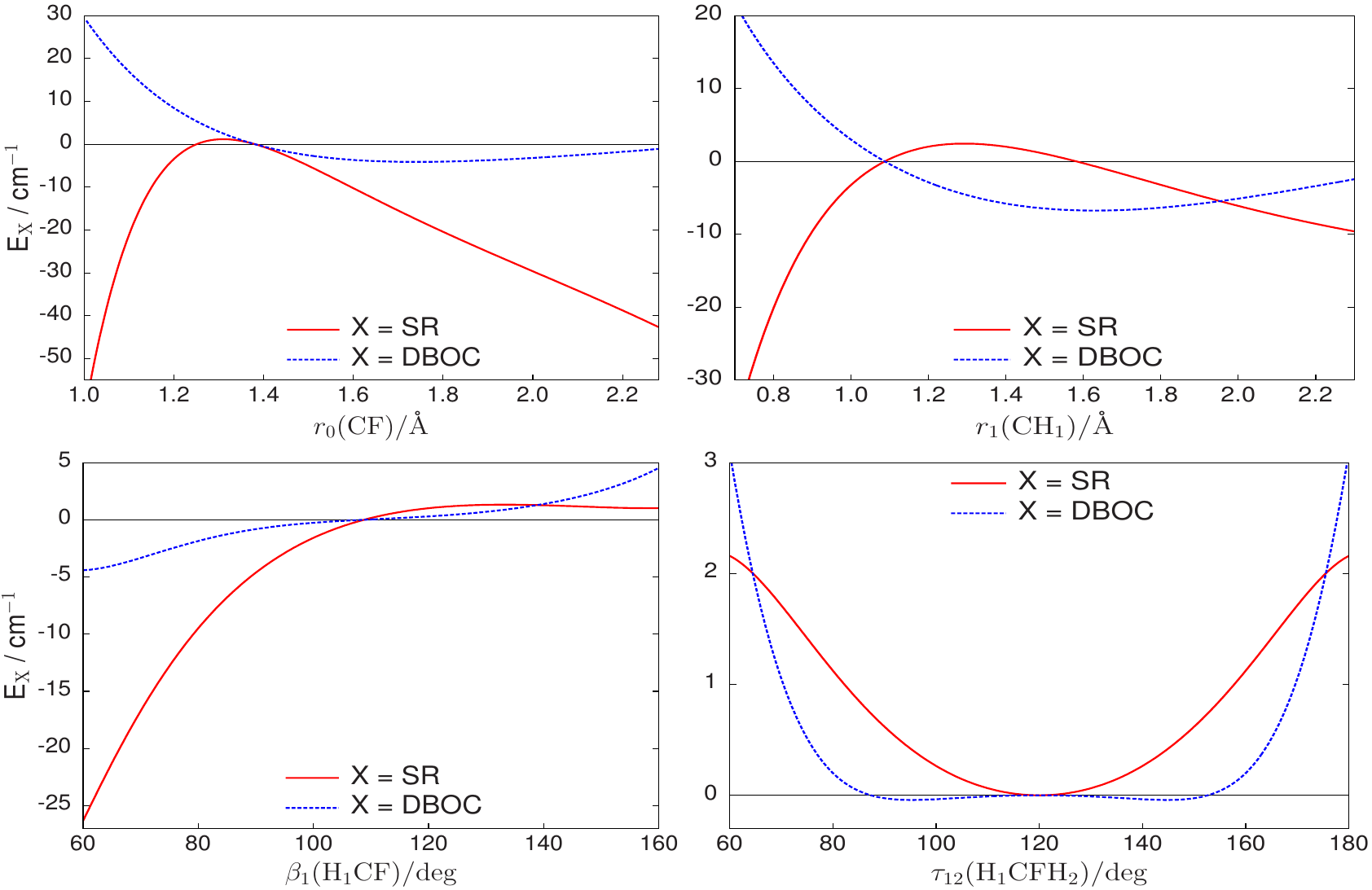}
\caption{\label{fig:1d_mvd1_dboc}One-dimensional cuts of the scalar relativistic (SR) and diagonal Born-Oppenheimer (DBOC) corrections with all other coordinates held at their equilibrium values.}
\end{figure}

 Computing the HL corrections at each grid point is computationally demanding but given the system size and chosen levels of theory is actually time-effective. Since the HL corrections vary in a smooth manner and are relatively small in magnitude, see Fig.~\ref{fig:1d_cv_ho} and Fig.~\ref{fig:1d_mvd1_dboc}, an alternative strategy would be to compute the HL corrections on reduced grids, fit suitable analytic representations to the data and then interpolate to other points on the global grid.~\cite{Yachmenev:JCP135:074302,Owens:JCP142:244306} This approach can be advantageous for larger systems or more computationally intensive electronic structure calculations. However, an adequate description of each HL correction requires careful consideration and is not necessarily straightforward. These issues are avoided in the present work.
  
\subsection{Analytic representation}

 Methyl fluoride is a prolate symmetric top molecule of $\bm{C}_{3\mathrm{v}}\mathrm{(M)}$ molecular symmetry.~\citep{Bunker:MolecularSymmetry} The \ce{XY3Z} symmetrized analytic representation utilized in this work has previously been employed for nuclear motion calculations of \ce{CH3Cl}.~\citep{Owens:JCP142:244306} Morse oscillator functions describe the stretching coordinates,
\begin{align}
\xi_1&=1-\exp\left[-a(r_0 - r_0^{\mathrm{ref}})\right] ,\label{eq:stretch1}\\
\xi_j&=1-\exp\left[-b(r_i - r_1^{\mathrm{ref}})\right]{\,};\hspace{2mm}j=2,3,4{\,}, \hspace{2mm} i=j-1 \label{eq:stretch2} ,
\end{align}
where $a=1.90$~{\AA}$^{-1}$ for the C{--}F internal coordinate $r_0$, and $b=1.87$~{\AA}$^{-1}$ for the three C{--}H internal coordinates $r_1,r_2$ and $r_3$. For the angular terms,
\begin{align}
\xi_k &= (\beta_i - \beta^{\mathrm{ref}}){\,};\hspace{2mm}k=5,6,7{\,}, \hspace{2mm} i=k-4 ,\label{eq:angular1}\\
\xi_8 &= \frac{1}{\sqrt{6}}\left(2\tau_{23}-\tau_{13}-\tau_{12}\right) ,\label{eq:angular2}\\
\xi_9 &= \frac{1}{\sqrt{2}}\left(\tau_{13}-\tau_{12}\right) ,\label{eq:angular3}
\end{align}
where $\tau_{23}=2\pi-\tau_{12}-\tau_{13}$, and $r_0^{\mathrm{ref}}$, $r_1^{\mathrm{ref}}$ and $\beta^{\mathrm{ref}}$ are the reference equilibrium structural parameters. Values of $r_0^{\mathrm{ref}}=1.3813$~{\AA}, $r_1^{\mathrm{ref}}= 1.0869$~{\AA}, and $\beta^{\mathrm{ref}}=108.773^{\circ}$ have been used, however, this choice is somewhat arbitrary due to the inclusion of linear expansion terms in the parameter set of the PES. Thus, the reference equilibrium structural parameters do not define the minimum of the PES and the true equilibrium values will be discussed in Sec.~\ref{sec:results_eq}.

 The potential function,
\begin{equation}\label{eq:pot_f}
V(\xi_{1},\xi_{2},\xi_{3},\xi_{4},\xi_{5},\xi_{6},\xi_{7},\xi_{8},\xi_{9})={\sum_{ijk\ldots}}{\,}\mathrm{f}_{ijk\ldots}V_{ijk\ldots} ,
\end{equation}
has maximum expansion order $i+j+k+l+m+n+p+q+r=6$ and is composed of the terms
\begin{equation}\label{eq:pot_term}
V_{ijk\ldots}=\lbrace\xi_{1}^{\,i}\xi_{2}^{\,j}\xi_{3}^{\,k}\xi_{4}^{\,l}\xi_{5}^{\,m}\xi_{6}^{\,n}\xi_{7}^{\,p}\xi_{8}^{\,q}\xi_{9}^{\,r}\rbrace^{\bm{C}_{3\mathrm{v}}\mathrm{(M)}} ,
\end{equation}
which are symmetrized combinations of different permutations of the coordinates $\xi_{i}$ that transform according to the $A_1$ representation of $\bm{C}_{3\mathrm{v}}\mathrm{(M)}$. The terms in Eq.~\eqref{eq:pot_term} are determined on-the-fly during the variational calculations and the PES implementation requires only a small amount of code.~\citep{Owens:JCP142:244306}

 To determine the expansion parameters $\mathrm{f}_{ijk\ldots}$ in Eq.~\eqref{eq:pot_f}, a least-squares fitting to the \textit{ab initio} data was carried out. Weight factors of the form proposed by \citet{Schwenke:JCP106:4618}
\begin{equation}\label{eq:weights}
w_i=\left(\frac{\tanh\left[-0.0006\times(\tilde{E}_i - 15{\,}000)\right]+1.002002002}{2.002002002}\right)\times\frac{1}{N\tilde{E}_i^{(w)}} ,
\end{equation}
were used in the fit. Here, $\tilde{E}_i^{(w)}=\max(\tilde{E}_i, 10{\,}000)$, where $\tilde{E}_i$ is the potential energy at the $i$th geometry above equilibrium and the normalization constant $N=0.0001$ (all units assume the energy is in cm$^{-1}$). Energies below 15,000~cm$^{-1}$ are favoured in our fitting by the weighting scheme. Watson's robust fitting scheme~\citep{Watson:JMolSpec219:326} was also utilized to further improve the description at lower energies and reduce the weights of outliers. The final PES was fitted with a weighted root-mean-square (rms) error of 0.97~cm$^{-1}$ for energies up to $h c \cdot 50{\,}000$~cm$^{-1}$ and required 405 expansion parameters.

 For geometries where $r_0\geq 1.95$~{\AA} and $r_i\geq 2.10$~{\AA} for $i=1,2,3$, the respective weights were reduced by several orders of magnitude. At larger stretch distances a T1 diagnostic value $>0.02$ indicates that the coupled cluster method has become unreliable.~\citep{Lee:IJQC36:199} Despite energies not being wholly accurate at these points, they are still useful and ensure that the PES maintains a reasonable shape towards dissociation. In subsequent calculations we refer to the PES as CBS-F12$^{\,\mathrm{HL}}$. The expansion parameters and a Fortran routine to construct the CBS-F12$^{\,\mathrm{HL}}$ PES are provided in the electronic supplementary information.

\section{Dipole moment surface}
\label{sec:DMS} 

\subsection{Electronic structure calculations}

 In a Cartesian laboratory-fixed $XYZ$ coordinate system with origin at the C nucleus, an external electric field with components $\pm0.005$~a.u. was applied along each coordinate axis and the respective dipole moment component $\mu_A$ for $A=X,Y,Z$ determined using central finite differences. Calculations were carried out in MOLPRO2012~\cite{Werner:WCMS2:242} at the CCSD(T)-F12b/aug-cc-pVTZ level of theory and employed the frozen core approximation with a Slater geminal exponent value of $\beta=1.2$~$a_0^{-1}$.~\citep{Hill:JCP131:194105} The same ansatz and ABS as in the explicitly correlated PES calculations were used. The DMS was computed on the same grid of nuclear geometries as the PES.

\subsection{Analytic representation}

 The analytic representation used for the DMS of methyl fluoride was previously employed for \ce{CH3Cl} and the reader is referred to \citet{Owens:JQSRT184:100} for a detailed description. To begin with, it is necessary to transform to a suitable molecule-fixed $xyz$ coordinate system before fitting an analytic expression to the \textit{ab initio} data. A unit vector is defined along each bond of \ce{CH3F},
\begin{equation}
\mathbf{e}_i = \frac{\mathbf{r}_i-\mathbf{r}_{\mathrm{C}}}{\lvert \mathbf{r}_i-\mathbf{r}_{\mathrm{C}}\rvert}{\,};\hspace{4mm}i=0,1,2,3 ,
\end{equation}
where $\mathbf{r}_{\mathrm{C}}$ is the position vector of the C nucleus, $\mathbf{r}_0$ the F nucleus, and $\mathbf{r}_1$, $\mathbf{r}_2$ and $\mathbf{r}_3$ the respective H atoms. The \textit{ab initio} dipole moment vector ${\bm\mu}$ is projected onto the molecular bonds and is described by three molecule-fixed $xyz$ dipole moment components,
\begin{align}
\mu_x &= \frac{1}{\sqrt{6}}\left[2({\bm\mu}\cdot\mathbf{e}_1) - ({\bm\mu}\cdot\mathbf{e}_2) - ({\bm\mu}\cdot\mathbf{e}_3) \right] , \label{eq:mu_x}\\
\mu_y &= \frac{1}{\sqrt{2}}\left[({\bm\mu}\cdot\mathbf{e}_2) - ({\bm\mu}\cdot\mathbf{e}_3) \right] , \label{eq:mu_y}\\
\mu_z &= {\bm\mu}\cdot\mathbf{e}_0 .\label{eq:mu_z}
\end{align}
We have formed symmetry-adapted combinations for $\mu_{x}$ and $\mu_{y}$ which transform according to the $E$ representation of $\bm{C}_{3\mathrm{v}}\mathrm{(M)}$, while the $\mu_{z}$ component is of $A_1$ symmetry. The symmetrized molecular bond representation described here is beneficial as the unit vectors $\mathbf{e}_i$ that define ${\bm\mu}$ for any instantaneous configuration of the nuclei are related to the internal coordinates only, meaning the description is self-contained.

 The three dipole moment surfaces $\mu_{\alpha}$ for $\alpha=x,y,z$ corresponding to Eqs.~\eqref{eq:mu_x} to \eqref{eq:mu_z} are represented by the analytic expression
\begin{equation}\label{eq:mu_tot}
\mu_{\alpha}(\xi_{1},\xi_{2},\xi_{3},\xi_{4},\xi_{5},\xi_{6},\xi_{7},\xi_{8},\xi_{9})={\sum_{ijk\ldots}}F^{(\alpha)}_{ijk\ldots}\mu_{\alpha,ijk\ldots}^{\Gamma=E,A_1} .
\end{equation}
The expansion terms
\begin{equation}\label{eq:mu_term}
\mu_{\alpha,ijk\ldots}^{\Gamma=E,A_1}=\lbrace\xi_{1}^{\,i}\xi_{2}^{\,j}\xi_{3}^{\,k}\xi_{4}^{\,l}\xi_{5}^{\,m}\xi_{6}^{\,n}\xi_{7}^{\,p}\xi_{8}^{\,q}\xi_{9}^{\,r}\rbrace^{\Gamma=E,A_1}_{\alpha},
\end{equation}
have maximum expansion order $i+j+k+l+m+n+p+q+r=6$ and are best understood as a sum of symmetrized combinations of different permutations of the coordinates $\xi_{i}$. Note that $\Gamma=E$ for $\mu_{x}$ and $\mu_{y}$, and $\Gamma=A_1$ for $\mu_{z}$. For the stretching coordinates we employed linear expansion variables,
\begin{align}
\xi_1&=\left(r_0 - r_0^{\mathrm{ref}}\right) ,\label{eq:stretch1_dms_ch3f}\\
\xi_j&=\left(r_i - r_1^{\mathrm{ref}}\right){\,};\hspace{2mm}j=2,3,4{\,}, \hspace{2mm} i=j-1 \label{eq:stretch2_dms_ch3f} ,
\end{align}
whilst the angular terms are the same as those defined in Eqs.~\eqref{eq:angular1} to \eqref{eq:angular3}. The reference structural parameters $r_0^{\mathrm{ref}}$, $r_1^{\mathrm{ref}}$ and $\beta^{\mathrm{ref}}$ had the same values as in the case of the PES.

 The expansion coefficients $F^{(\alpha)}_{ijk\ldots}$ for $\alpha=x,y,z$ were determined simultaneously through a least-squares fitting to the \textit{ab initio} data. Weight factors of the form given in Eq.~\eqref{eq:weights} were used along with Watson's robust fitting scheme.~\citep{Watson:JMolSpec219:326} The three dipole moment surfaces, $\mu_x$, $\mu_y$, and $\mu_z$, required 171, 160 and 226 parameters, respectively. A combined weighted rms error of $1\times10^{-4}$~D was achieved for the fitting. Similar to the PES, the analytic representation of the DMS is generated on-the-fly at runtime. Its construction is slightly more complex because ${\bm\mu}$ is a vector quantity and the transformation properties of the dipole moment components must also be considered.~\citep{Owens:JQSRT184:100} The expansion parameter set of the DMS is given in the electronic supplementary information along with a Fortran routine to construct the analytic representation.

\section{Variational calculations}
\label{sec:variational} 
 
 The general methodology of {\sc TROVE} is well documented~\citep{Yurchenko:JMS245:126,Yurchenko:JPCA113:11845,Yachmenev:JCP143:014105,Yurchenko:JCTC13:4368} and calculations on another \ce{XY3Z} molecule, namely \ce{CH3Cl}, have previously been reported.~\citep{Owens:JCP142:244306,Owens:JQSRT184:100} We therefore summarize only the key aspects relevant for this work.
 
 The rovibrational Hamiltonian was represented as a power series expansion around the equilibrium geometry in terms of the nine coordinates introduced in Eqs.~\eqref{eq:stretch1} to \eqref{eq:angular3}. However, for the kinetic energy operator linear displacement variables $(r_i - r_{\mathrm{ref}})$ were used for the stretching coordinates. The Hamiltonian was constructed numerically using an automatic differentiation method~\cite{Yachmenev:JCP143:014105} with both the kinetic and potential energy operators truncated at sixth order. The associated errors of such a scheme are discussed in \citet{Yurchenko:JMS245:126} and \citet{Yachmenev:JCP143:014105}. Atomic mass values~\citep{IUPAC:Cohen} were employed throughout.
 
\begin{figure}
\centering
\includegraphics{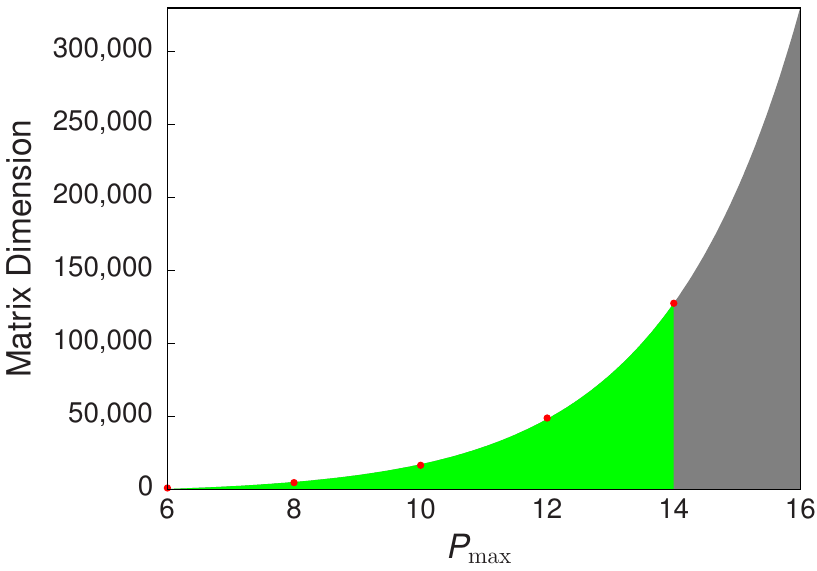}
\caption{\label{fig:dimension}Size of the $J\!=\!0$ Hamiltonian matrix with respect to the polyad truncation number $P_{\mathrm{max}}$. Calculations were not possible above $P_{\mathrm{max}}=14$.}
\end{figure}

 A multi-step contraction scheme~\cite{Yurchenko:JCTC13:4368} was used to construct the symmetrized vibrational basis set, the size of which was controlled by the polyad number,
\begin{equation}\label{eq:polyad}
P = n_1+2(n_2+n_3+n_4)+n_5+n_6+n_7+n_8+n_9 \leq P_{\mathrm{max}} ,
\end{equation}
and this does not exceed a predefined maximum value $P_{\mathrm{max}}$. Here, the quantum numbers $n_k$ for $k=1,\ldots,9$ relate to primitive basis functions $\phi_{n_k}$, which are obtained by solving one-dimensional Schr\"{o}dinger equations for each $k$th vibrational mode using the Numerov-Cooley method.~\cite{Numerov:MNRAS84:592,Cooley:MathComp15:363} Multiplication with symmetrized rigid-rotor eigenfunctions $\ket{J,K,m,\tau_{\mathrm{rot}}}$ produced the final basis set for use in $J>0$ calculations. The quantum numbers $K$ and $m$ are the projections (in units of $\hbar$) of $J$ onto the molecule-fixed $z$ axis and the laboratory-fixed $Z$ axis, respectively, whilst $\tau_{\mathrm{rot}}$ determines the rotational parity as $(-1)^{\tau_{\mathrm{rot}}}$. As shown in Fig.~\ref{fig:dimension}, the size of the Hamiltonian matrix, i.e., the number of $J=0$ basis functions, grows exponentially with respect to $P_{\mathrm{max}}$ and {\sc TROVE} calculations above $P_{\mathrm{max}}=14$ were not possible with the resources available to us. We will see in Sec.~\ref{sec:results} that differently sized basis sets and basis set techniques must be utilized when computing various spectroscopic quantities due to the computational demands of variational calculations.

\section{Results}
\label{sec:results}

\subsection{Equilibrium geometry and pure rotational energies}
\label{sec:results_eq}

 The equilibrium geometry derived from the CBS-F12$^{\,\mathrm{HL}}$ PES is listed in Table~\ref{tab:eq_ref}. It is in excellent agreement with previous values determined in a joint experimental and \textit{ab initio} analysis by \citet{Demaison:StructChem10:129}, which is regarded as the most reliable equilibrium structure of \ce{CH3F} to date. Also validating is the agreement with the refined geometry PES of \citet{Nikitin:JMolSpec274:28}, particularly as the CBS-F12$^{\,\mathrm{HL}}$ PES has been generated in a purely \textit{ab initio} fashion.
 
\begin{table}
\footnotesize
\tabcolsep=0.06cm
\caption{\label{tab:eq_ref}The equilibrium structural parameters of \ce{CH3F}}
\begin{center}
\begin{tabular}{l l l l l}
\hline
$r$(C{--}F) / {\AA} & $r$(C{--}H) / {\AA} & $\beta$(HCF) / deg  & Reference & Approach\\
\hline
1.38242 & 1.08698 & 108.746 & This work & Purely \textit{ab initio} PES\\
1.3826(3) & 1.0872(3) & 108.69(4) & \citet{Demaison:StructChem10:129} & Experimental and \textit{ab initio} analysis\\
1.3827 & 1.0876 & 108.75 & \citet{Demaison:StructChem10:129} & \textit{Ab initio} calculations\\
1.38240 & 1.08696 & 108.767 & \citet{Nikitin:JMolSpec274:28} & Refined geometry PES\\
\hline
\end{tabular}
\end{center}
\end{table}

 It is more illustrative to look at the pure rotational energy levels, shown in Table~\ref{tab:rotational_ch3f}, since these are highly dependent on the molecular geometry through the moments of inertia. Calculations with {\sc TROVE} employed a polyad truncation number of $P_{\mathrm{max}}=8$, which is sufficient for converging ground state rotational energies. Despite being consistently lower than the experimental values, the rotational energies up to $J\leq5$ are reproduced with an rms error of 0.0015~cm$^{-1}$. The residual error $\Delta E(\mathrm{obs}-\mathrm{calc})$ increases at each step up in $J$, however, this can be easily counteracted by refining the equilibrium geometry of the CBS-F12$^{\,\mathrm{HL}}$ PES through a nonlinear least-squares fitting to the experimental energies, for example, see \citet{Owens:JCP145:104305}. The accuracy of the computed intra-band rotational wavenumbers can be substantially improved as a result, but we refrain from doing this here as the errors for \ce{CH3F} are very minor and it leads to a poorer description of the vibrational energy levels.

\begin{table}[!ht]
\tabcolsep=0.35cm
\caption{\label{tab:rotational_ch3f}Comparison of computed and experimental $J\leq5$ pure rotational term values (in cm$^{-1}$) for \ce{CH3F}. The observed ground state energy levels are taken from \citet{Nikitin:JMolSpec274:28} but are attributed to \citet{Demaison:StructChem10:129}}
\begin{center}
\begin{tabular}{cccrrr}
\hline
 $J$ & $K$ & Sym. & Experiment & Calculated & Obs$-$calc \\
\hline
 0&  0& $A_1$ &    0.00000&     0.00000&  0.00000\\
 1&  0& $A_2$ &    1.70358&     1.70348&  0.00010\\
 1&  1& $E$   &    6.03369&     6.03352&  0.00017\\
 2&  0& $A_1$ &    5.11069&     5.11040&  0.00029\\
 2&  1& $E$   &    9.44074&     9.44038&  0.00036\\
 2&  2& $E$   &   22.43005&    22.42946&  0.00059\\
 3&  0& $A_2$ &   10.22124&    10.22066&  0.00058\\
 3&  1& $E$   &   14.55120&    14.55055&  0.00065\\
 3&  2& $E$   &   27.54025&    27.53937&  0.00088\\
 3&  3& $A_1$ &   49.18585&    49.18459&  0.00126\\
 3&  3& $A_2$ &   49.18585&    49.18459&  0.00126\\
 4&  0& $A_1$ &   17.03508&    17.03411&  0.00097\\
 4&  1& $E$   &   21.36493&    21.36388&  0.00105\\
 4&  2& $E$   &   34.35362&    34.35235&  0.00127\\
 4&  3& $A_1$ &   55.99863&    55.99698&  0.00165\\
 4&  3& $A_2$ &   55.99863&    55.99698&  0.00165\\
 4&  4& $E$   &   86.29575&    86.29356&  0.00219\\
 5&  0& $A_2$ &   25.55202&    25.55057&  0.00145\\
 5&  1& $E$   &   29.88172&    29.88019&  0.00153\\
 5&  2& $E$   &   42.86997&    42.86822&  0.00175\\
 5&  3& $A_1$ &   64.51425&    64.51212&  0.00213\\
 5&  3& $A_2$ &   64.51425&    64.51212&  0.00213\\
 5&  4& $E$   &   94.81034&    94.80767&  0.00267\\
 5&  5& $E$   &  133.75234&   133.74897&  0.00337\\
\hline
\end{tabular}
\end{center}
\end{table}

\subsection{Vibrational $J=0$ energy levels}

 To assess the CBS-F12$^{\,\mathrm{HL}}$ PES it is necessary to have converged vibrational energy levels, and one method of ensuring this is a complete vibrational basis set (CVBS) extrapolation.~\cite{Ovsyannikov:JCP129:044309} Similar to basis set extrapolation techniques of electronic structure theory (see e.g. \citet{Feller:JCP138:074103} and references therein), the same ideas can be applied to {\sc TROVE} calculations with respect to $P_{\mathrm{max}}$. Vibrational energies were computed for $P_{\mathrm{max}}=\lbrace 10,12,14 \rbrace$ and fitted using the exponential decay expression,
\begin{equation}
 E_i(P_{\mathrm{max}}) = E_i^{\mathrm{CVBS}}+A_i\exp(-\lambda_i P_{\mathrm{max}}) ,
\end{equation}
where $E_i$ is the energy of the $i$th level, $E_i^{\mathrm{CVBS}}$ is the respective energy at the CVBS limit, $A_i$ is a fitting parameter, and $\lambda_i$ is determined from
\begin{equation}
\lambda_i=-\frac{1}{2}\ln\left(\frac{E_i(P_{\mathrm{max}}\!=\!14)-E_i(P_{\mathrm{max}}\!=\!12)}{E_i(P_{\mathrm{max}}\!=\!12)-E_i(P_{\mathrm{max}}\!=\!10)}\right) .
\end{equation}

 The CVBS extrapolated $J=0$ energies are shown in Table~\ref{tab:j0_ch3f} alongside known experimental values.~\cite{Papousek:JMolSpec196:319,Papousek:JMolSpec149:109,Betrencourt:MP33:83,Champion:JMolSpec96:422,Graner:JPC83:1491,Kondo:JPC90:1519,
 Blass:JMolSpec25:440,Jones:PRSLA290:490} The six fundamentals are reproduced with an rms error of 0.69~cm$^{-1}$ and a mean-absolute-deviation (mad) of 0.53~cm$^{-1}$. This level of accuracy extends to most of the other term values which are well within the $\pm$1~cm$^{-1}$ accuracy expected from PESs based on high-level \textit{ab initio} theory. Most significant perhaps is the computed $\nu_4$ level which shows a residual error $\Delta E(\mathrm{obs}-\mathrm{calc})$ of $-$1.26~cm$^{-1}$. This is a noticeable improvement compared to the PES of \citet{Zhao:JCP144:204302} ($\Delta E_{\nu_4}$=3.33~cm$^{-1}$), which was generated at the CCSD(T)-F12a/aug-cc-pVTZ level of theory, and the PES of \citet{Nikitin:JMolSpec274:28} ($\Delta E_{\nu_4}$=4.86~cm$^{-1}$), computed at the CCSD(T)/cc-pVQZ level of theory with relativistic corrections, thus highlighting the importance of including HL corrections and a CBS extrapolation in the PES of \ce{CH3F}.
 
 It is worth noting, at least for the values considered in Table~\ref{tab:j0_ch3f}, that the computed $P_{\mathrm{max}}=14$ vibrational energy levels are within 0.01~cm$^{-1}$ of the CVBS values with the majority converged to one or two orders-of-magnitude better. A complete list of the $P_{\mathrm{max}}=14$ computed $J=0$ energy levels is included in the electronic supplementary information.

\begin{table}[!ht]
\tabcolsep=0.35cm
\caption{\label{tab:j0_ch3f}Comparison of computed and experimental $J=0$ vibrational term values (in cm$^{-1}$) for \ce{CH3F}. The computed zero-point energy was 8560.2409~cm$^{-1}$ at the CVBS limit}
\begin{center}
\begin{tabular}{lccccc}
\hline
Mode & Sym. &  Experiment & Calculated & Obs$-$calc & Ref. \\
\hline
$\nu_3$       & $A_1$ & 1048.61& 1048.88& $-$0.27& \citenum{Papousek:JMolSpec196:319} \\
$\nu_6$       & $E$   & 1182.67& 1182.79& $-$0.12& \citenum{Papousek:JMolSpec196:319} \\
$\nu_2$       & $A_1$ & 1459.39& 1459.67& $-$0.28& \citenum{Papousek:JMolSpec196:319} \\
$\nu_5$       & $E$   & 1467.81& 1468.03& $-$0.21& \citenum{Papousek:JMolSpec196:319} \\
$2\nu_3$      & $A_1$ & 2081.38& 2081.81& $-$0.43& \citenum{Papousek:JMolSpec149:109} \\
$\nu_3+\nu_6$ & $E$   & 2221.81& 2222.20& $-$0.40& \citenum{Papousek:JMolSpec149:109} \\
$2\nu_6$      & $E$   & 2365.80& 2365.96& $-$0.16& \citenum{Betrencourt:MP33:83} \\
$\nu_2+\nu_3$ & $A_1$ & 2499.80& 2500.28& $-$0.48& \citenum{Betrencourt:MP33:83} \\
$\nu_3+\nu_5$ & $E$   & 2513.80& 2514.27& $-$0.47& \citenum{Betrencourt:MP33:83} \\
$2\nu_5$      & $A_1$ & 2863.24& 2863.90& $-$0.66& \citenum{Champion:JMolSpec96:422} \\
$\nu_2+\nu_5$ & $E$   & 2922.23& 2922.59& $-$0.36& \citenum{Champion:JMolSpec96:422} \\
$2\nu_2$      & $A_1$ & 2926.00& 2926.66& $-$0.66& \citenum{Champion:JMolSpec96:422} \\
$2\nu_5$      & $E$   & 2927.39& 2927.92& $-$0.53& \citenum{Champion:JMolSpec96:422} \\
$\nu_1$       & $A_1$ & 2966.25& 2967.30& $-$1.05& \citenum{Champion:JMolSpec96:422} \\
$\nu_4$       & $E$   & 3005.81& 3007.07& $-$1.26& \citenum{Champion:JMolSpec96:422} \\
$3\nu_3$      & $A_1$ & 3098.44& 3098.97& $-$0.53& \citenum{Graner:JPC83:1491} \\
$\nu_3+2\nu_5$& $A_1$ &  3905.4& 3906.39& $-$0.99& \citenum{Kondo:JPC90:1519} \\
$\nu_1+\nu_3$ & $A_1$ &    4011& 4012.28& $-$1.28& \citenum{Blass:JMolSpec25:440} \\
$\nu_3+\nu_4$ & $E$   &  4057.6& 4059.31& $-$1.71& \citenum{Jones:PRSLA290:490} \\
$2\nu_4$      & $E$   & 6000.78& 6003.11& $-$2.33& \citenum{Jones:PRSLA290:490} \\
\hline
\end{tabular}
\end{center}
\end{table}

\subsection{Vibrational transition moments}

 The vibrational transition moment is defined as,
\begin{equation}
\label{eq:TM}
\mu_{if} = \sqrt{\sum_{\alpha=x,y,z}{\lvert\bra{\Phi^{(f)}_{\mathrm{vib}}}\bar{\mu}_{\alpha}\ket{\Phi^{(i)}_{\mathrm{vib}}}\rvert}^2} ,
\end{equation}
where $\ket{\Phi^{(i)}_{\mathrm{vib}}}$ and $\ket{\Phi^{(f)}_{\mathrm{vib}}}$ are the $J=0$ initial and final state eigenfunctions, respectively, and $\bar{\mu}_{\alpha}$ is the electronically averaged dipole moment function along the molecule-fixed axis $\alpha=x,y,z$. Transition moments are relatively inexpensive to compute and provide an initial assessment of the DMS. Calculations in {\sc TROVE} used $P_{\mathrm{max}}=12$ and considered transitions from the ground vibrational state only ($i=0$).

 In Table~\ref{tab:tm_ch3f}, vibrational transition moments for the fundamentals are listed alongside known experimental values derived from measurements of absolute line intensities.~\cite{Lepere:JMolSpec177:307,Lepere:JMolSpec180:218,Lepere:JMolSpec189:137} The agreement is encouraging but it suggests that the DMS may overestimate the strength of line intensities. This behaviour will be confirmed in the following sections. A list of computed transition moments from the vibrational ground state up to 10,000~cm$^{-1}$ is provided in the electronic supplementary information. For the ground-state electric dipole moment of \ce{CH3F} we compute a value of 1.8503~D, which is close to the experimentally determined value of 1.8584~D.~\cite{Marshall:JMolSpec83:279}
 
\begin{table}[!ht]
\tabcolsep=0.35cm
\caption{\label{tab:tm_ch3f}Vibrational transition moments (in Debye) for the fundamental frequencies (in cm$^{-1}$) of \ce{CH3F}}
\begin{center}
\begin{tabular}{cccccc}
\hline
Mode & Sym. & $\nu_{0f}^{\mathrm{exp}}$ & \multicolumn{1}{c}{$\mu_{0f}^{\mathrm{calc}}$} & \multicolumn{1}{c}{$\mu_{0f}^{\mathrm{exp}}$} & \multicolumn{1}{c}{Ref.}\\
\hline
$\nu_1$ & $A_1$ & 2966.25 & 0.05205 & -- & --\\
$\nu_2$ & $A_1$ & 1459.39 & 0.01390 & 0.01196 & \citenum{Lepere:JMolSpec189:137}\\
$\nu_3$ & $A_1$ & 1048.61 & 0.20020 & 0.19015 & \citenum{Lepere:JMolSpec177:307}\\
$\nu_4$ & $E$ & 3005.81 & 0.08485 & -- & --\\
$\nu_5$ & $E$ & 1467.81 & 0.04903 & 0.04976 & \citenum{Lepere:JMolSpec189:137}\\
$\nu_6$ & $E$ & 1182.67 & 0.03085 & 0.02835& \citenum{Lepere:JMolSpec180:218}\\
\hline
\end{tabular}
\end{center}
\end{table}

\subsection{Absolute line intensities of the $\nu_6$ band}

 Recently, \citet{Jacquemart:JQSRT185:58} generated an experimental line list of almost 1500 transitions of the $\nu_6$ band with absolute line intensities determined with an estimated accuracy of 5\%. To compare with this study we have generated an \textit{ab initio} room temperature line list for \ce{CH3F}. This was computed with a lower state energy threshold of $h c \cdot$~5000~cm$^{-1}$ and considered transitions up to $J=40$ in the 0--4600~cm$^{-1}$ range. 
 
 Describing high rotational excitation can quickly become computationally intractable since rovibrational matrices scale linearly with $J$. It was therefore necessary to use a truncated $J=0$ basis set. {\sc TROVE} calculations were initially performed with $P_{\mathrm{max}}=12$, resulting in 49,076 vibrational basis functions, which was subsequently reduced to 2153 functions by removing states with energies above $h c \cdot$~9600~cm$^{-1}$. The resulting pruned basis set was multiplied in the usual manner with symmetrized rigid-rotor functions to produce the final basis set for $J>0$ calculations. 
 
 Naturally, errors are introduced into our rovibrational predictions and it is hard to quantify this without more rigorous calculations. However, we have previously used truncated basis set procedures to construct a comprehensive line list for \ce{SiH4} without noticeable deterioration.~\cite{Owens:MNRAS471:5025} It should be emphasised that the main advantage of truncation is the ability to retain the accuracy of the vibrational energy levels and respective wavefunctions generated with $P_{\mathrm{max}}=12$.

 Absolute absorption intensities were simulated using the expression,
\begin{equation}
\label{eq:abs_I}
I(f \leftarrow i) = \frac{A_{if}}{8\pi c}g_{\mathrm{ns}}(2 J_{f}+1)\frac{\exp\left(-E_{i}/kT\right)}{Q(T)\; \nu_{if}^{2}}\left[1-\exp\left(-\frac{hc\nu_{if}}{kT}\right)\right] ,
\end{equation}
where $A_{if}$ is the Einstein-$A$ coefficient of a transition with wavenumber $\nu_{if}$ (in cm$^{-1}$) between an initial state with energy $E_i$ and a final state with rotational quantum number $J_f$. Here, $k$ is the Boltzmann constant, $h$ is the Planck constant, $c$ is the speed of light and the absolute temperature $T=296$~K. The nuclear spin statistical weights are $g_{\mathrm{ns}}=\lbrace 8,8,8\rbrace$ for states of symmetry $\lbrace A_1,A_2,E\rbrace$, respectively, and for the room temperature partition function $Q(296\,K)=$~ 14,587.780.~\cite{Rey:JMolSpec327:138} Transitions obey the symmetry selection rules $A_1 \leftrightarrow A_2,\; E \leftrightarrow E$; and the standard rotational selection rules, $J\p-J\pp=0,\pm 1,\; J\p+J\pp \ne 0$; where $\p$ and $\pp$ denote the upper and lower state, respectively. The \textsc{ExoCross} code~\cite{Yurchenko:AA:inpress} was employed for all spectral simulations.

\begin{figure}
\centering
\includegraphics[width=\textwidth]{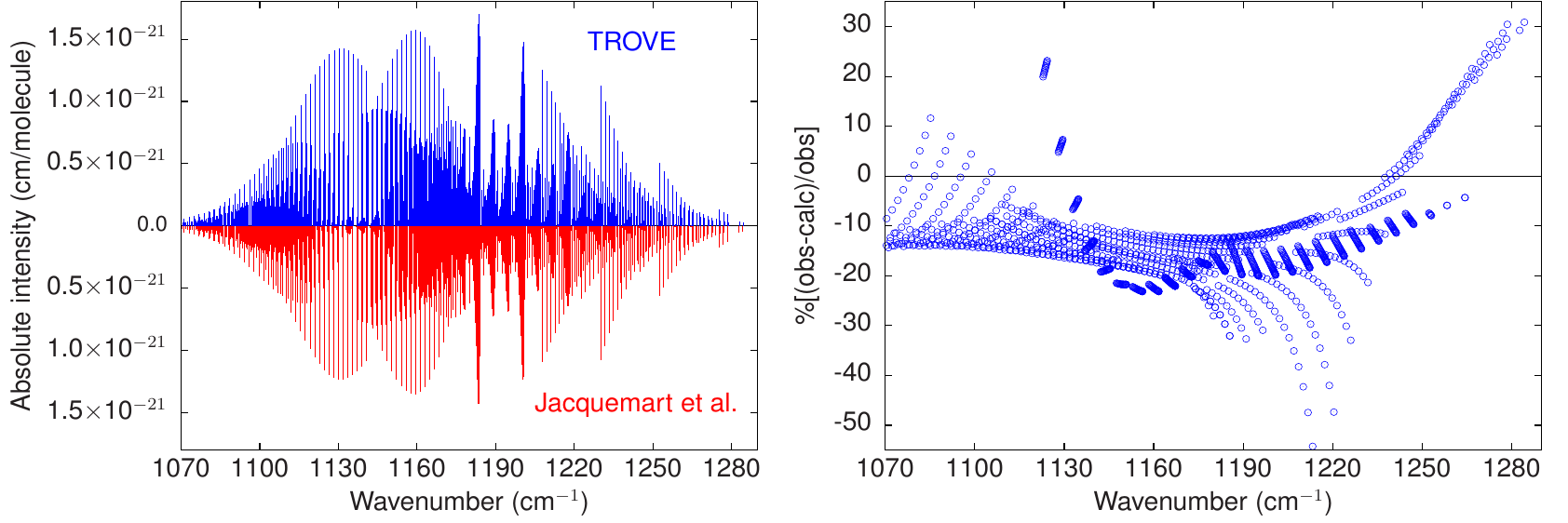}
\caption{\label{fig:v6}Absolute line intensities of the $\nu_6$ band for transitions up to $J=40$ at $T=$~296~K and the corresponding residual errors $\%[\mathrm{(obs-calc)/obs}]$ when compared with the experimental line list from \citet{Jacquemart:JQSRT185:58}.}
\end{figure}

 In Fig.~\ref{fig:v6}, the computed absolute line intensities of the $\nu_6$ band are plotted against the experimental line list of \citet{Jacquemart:JQSRT185:58} alongside the percentage measure $\%[\mathrm{(obs-calc)/obs}]$, which quantifies the error in our predicted intensities. The shape and structure of the $\nu_6$ band is well reproduced but the DMS overestimates the strength of line intensities. We expect that this behaviour can be corrected for by using a larger augmented basis set in the electronic structure calculations, however, the improvement in intensities may not justify the additional computational expense. As expected from the $J=0$ calculations, computed line positions of the $\nu_6$ band had an average residual error of $\Delta \nu(\mathrm{obs}-\mathrm{calc})=-$0.125~cm$^{-1}$.

\subsection{Overview of the rotation-vibration spectrum}

\begin{figure}
\centering
\includegraphics[width=\textwidth]{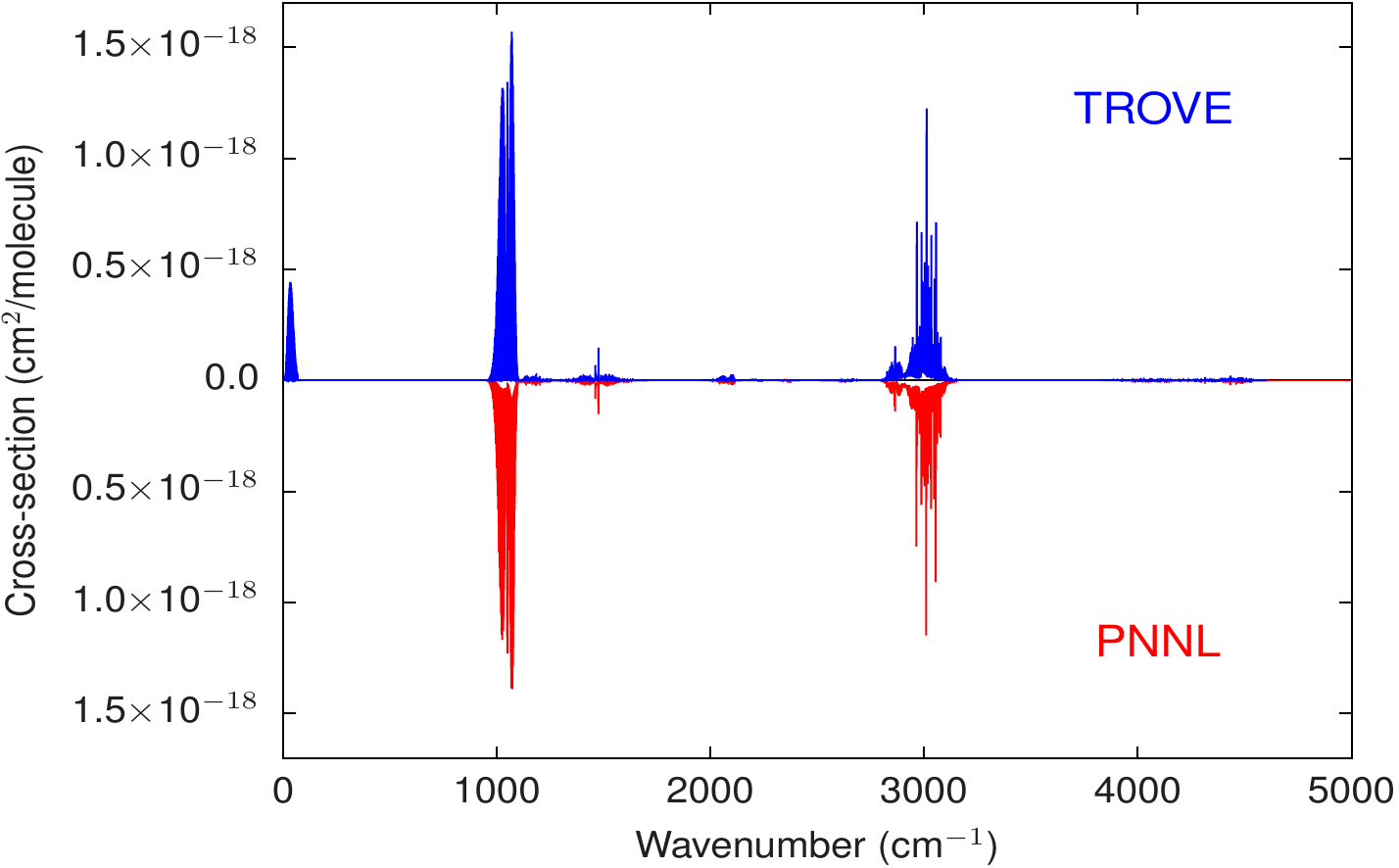}
\caption{\label{fig:pnnl_whole}Simulated \ce{CH3F}  rotation-vibration spectrum up to $J=40$ compared with the PNNL spectral library~\cite{Sharpe:AppSpec58:1452} at $T=$~296~K.}
\end{figure}

\begin{figure}
\centering
\includegraphics[width=\textwidth]{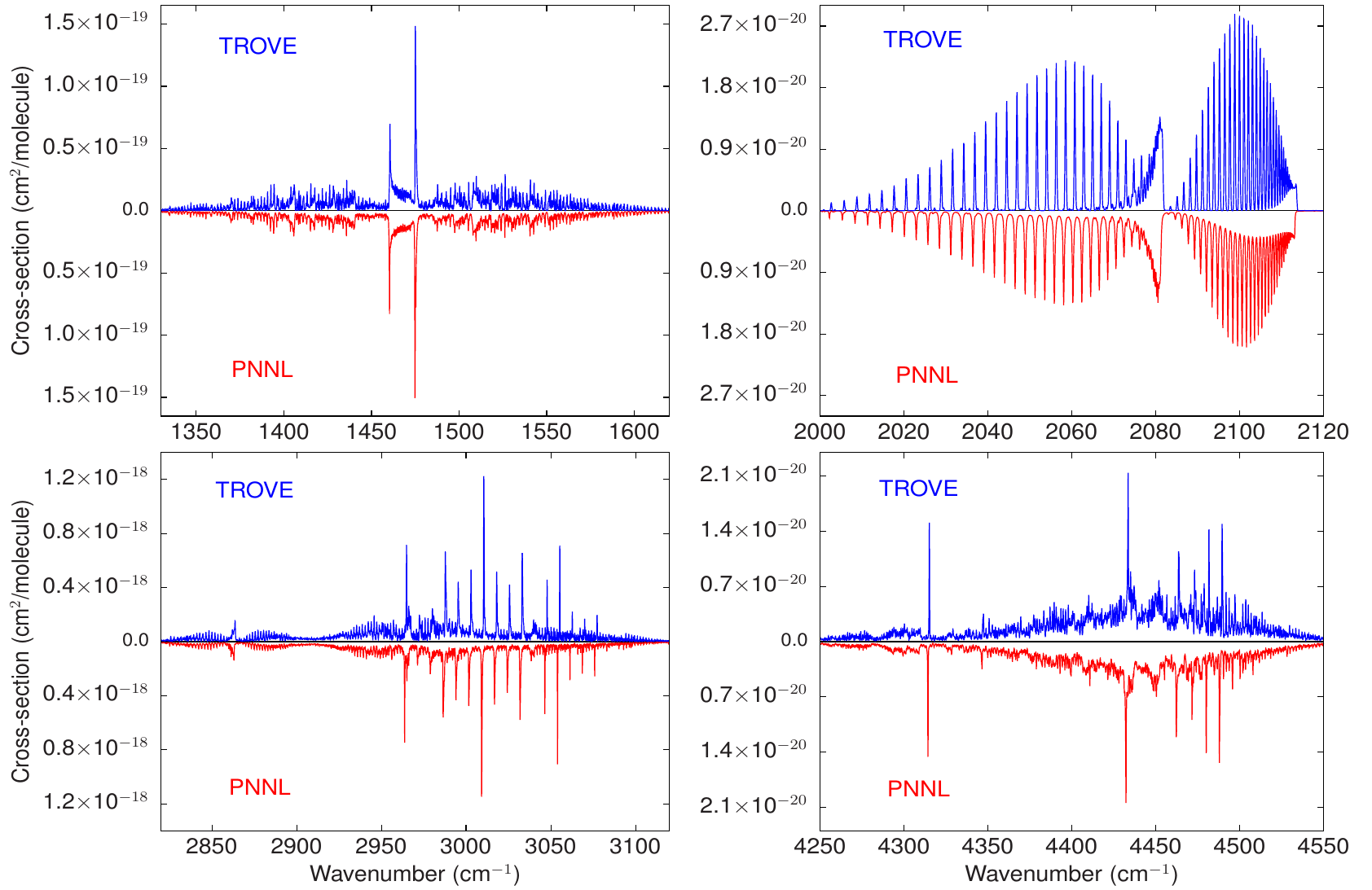}
\caption{\label{fig:pnnl_panels}Closer inspection of the computed cross-sections compared with the PNNL spectral library~\cite{Sharpe:AppSpec58:1452} at $T=$~296~K.}
\end{figure}

 A final benchmark of our rovibrational calculations and \textit{ab initio} line list is a comparison with the PNNL spectral library.~\citep{Sharpe:AppSpec58:1452} Absorption cross-sections were simulated at a resolution of 0.06~cm$^{-1}$ using a Gaussian profile with a half-width at half-maximum of $0.135$~cm$^{-1}$. The results are shown in Fig.~\ref{fig:pnnl_whole} and Fig.~\ref{fig:pnnl_panels} with the experimental PNNL spectrum, which was measured at a temperature of 25$\,^{\circ}$C with the dataset subsequently re-normalized to 22.84$\,^{\circ}$C ($296$~K). Overall the agreement is extremely pleasing, particularly as both the strong and weak intensity features are equally well reproduced. Whilst the intensities of the \textit{ab initio} spectrum are stronger, this is only slight and we have no hesitation recommending the PES and DMS for future use in spectroscopic applications.

\section{Conclusions}
\label{sec:conc}

 A new PES and DMS for methyl fluoride have been generated using high-level \textit{ab initio} theory and then rigorously evaluated through variational nuclear motion calculations. The computed results showed excellent agreement with a range of experimental data, which included the equilibrium geometry of \ce{CH3F}, pure rotational and vibrational energies, vibrational transition moments, absolute line intensities of the $\nu_6$ band, and the rotation-vibration spectrum up to $J=40$. This work demonstrates the importance of including HL energy corrections and an extrapolation to the CBS limit in the PES to accurately describe the rovibrational spectrum of \ce{CH3F} from first principles.
 
 To go beyond the accuracy achieved in this work in a purely \textit{ab initio} manner will require extensive larger basis set electronic structure calculations. That said, the computational cost associated with this is unlikely to correlate with the gain in accuracy and empirical refinement of the PES is recommended instead. Although computationally intensive,~\cite{Yurchenko:JMolSpec268:123} refinement can lead to orders-of-magnitude improvements in the accuracy of the computed rovibrational energy levels and consequently more reliable transition intensities as a result of better wavefunctions.

\section*{Supplementary Material}

Electronic Supplementary Information (ESI) available: The expansion parameters and Fortran 90 functions to construct the PES and DMS of \ce{CH3F}. A list of computed vibrational energy levels and vibrational transition moments.

\begin{acknowledgments}
Besides DESY, this work has been supported by the \emph{Deutsche Forschungsgemeinschaft} (DFG) through the excellence cluster ``The Hamburg Center for Ultrafast Imaging -- Structure, Dynamics and Control of Matter at the Atomic Scale'' (CUI, EXC1074) and the priority program 1840 ``Quantum Dynamics in Tailored Intense Fields'' (QUTIF, KU1527/3), by the Helmholtz Association ``Initiative and Networking Fund'', and by the COST action MOLIM (CM1405).
\end{acknowledgments}

\clearpage
\newpage
%

\end{document}